\documentclass[12pt,a4paper]{article}
\usepackage{amsmath}
\usepackage{amssymb} 
\usepackage{graphicx} 
\usepackage{float}
\restylefloat{table}
\begin{document}
\begin{titlepage} 
\begin{flushright} IFUP--TH/2014-1\\ 
\end{flushright} ~
\vskip .8truecm 
\begin{center} 
\Large\bf On the monodromy problem for the four-punctured sphere
\end{center}
\vskip 1.2truecm 
\begin{center}
{Pietro Menotti} \\ 
{\small\it Dipartimento di Fisica, Universit{\`a} di Pisa}\\ 
{\small\it 
Largo B. Pontecorvo 3, I-56127, Pisa, Italy}\\
{\small\it e-mail: menotti@df.unipi.it}\\ 
\end{center} 
\vskip 0.8truecm
\centerline{April 2014}
                
\vskip 1.2truecm
                                                              
\begin{abstract}
We consider the monodromy problem for the four-punctured sphere in
which the character of one composite monodromy is fixed, by looking at
the expansion of the accessory parameter in the modulus $x$ directly, 
without taking the limit of the
quantum conformal blocks for infinite central charge. The integrals
which appear in the expansion of the Volterra equation, involve
products of two hypergeometric functions to first order and up to four
hypergeometric functions to second order. It is shown that all such
integrals can be computed analytically. We give the complete
analytical evaluation of the accessory parameter to first and second
order in the modulus. The results agree with the evaluation obtained
by assuming the exponentiation hypothesis of the quantum conformal
blocks in the limit of infinite central charge. Extension to
higher orders is discussed.
\end{abstract}

\end{titlepage}
 
\eject

\section{Introduction}

In the papers \cite{LLNZ,LRS} the following monodromy problem is considered
for the Liouville theory on the sphere: Given the singularities in the
standard position $0,x,1,\infty$ and given the class i.e. the trace of
the monodromy for a path encircling both singularities at $0$ and $x$,
find the value of the accessory parameter realizing such data.  
Such a problem intervenes in the process of the classical limit of the
quantum four-point function; the value of trace of the above described
monodromy is then fixed by a saddle point procedure \cite{ZZ}. 
In \cite{ZZ,LLNZ} the problem is solved by going over to the quantum 
formulation for
the four-point function and taking the classical limit i.e. 
the limit in which the central charge goes to infinity.
As the quantum conformal blocks are known as
formal power series expansions in $x$ also the classical result so 
obtained is
given as a formal power expansion in $x$. The procedure goes through a
process of exponentiation of the quantum conformal blocks after which
the classical limit $b\rightarrow 0$ is taken; 
in such a limit heavy cancellations take part \cite{hadaszjaskolski1}.
Several analytic \cite{hadaszjaskolski2,ferraripiatek,LLNZ,poghossian} 
and numerical \cite{ZZ,hadaszjaskolski1} calculations, also exploiting
recursion formulae \cite{AlZ1,AlZ2} for the conformal blocks, support
the validity of such a calculational scheme.

One suspects, on the other hand, that the same results should be
obtainable just by exploiting the transformation properties of the
ordinary differential equations which underlie the Liouville theory
at the classical level. In this note we shall in fact consider the
problem directly at the classical level.
In addition it appears that working without taking the singular limit
in which the central charge goes to infinity one might control better
the convergence region of the expansion of the accessory parameter as
a function of the modulus \cite{torusIII}.

The approach followed in this paper consists in computing the monodromy 
along a contour embracing $0$ and $x$ through the usual convergent 
iteration expansion for the solution of the Heun equation. The
monodromy is computed along a contour which avoids the neighborhood of 
the origin where the kernel is singular and then we expand
the result in $x$.

In so doing one is faced to first order with the computation of
integrals containing
the product of two hypergeometric functions; if one goes to the second
order the product of four hypergeometric functions in a double integral
appears. 

In this paper we
show how to compute analytically such integrals, which appear in the
expansion of the solution of the Volterra equation. The
complete first order result gives as a byproduct the value of the
accessory parameter which coincides with the one derived in \cite{ZZ}
and re-derived in \cite{ferraripiatek,LLNZ}. 

For computing the integrals appearing in the first order result, we
exploit the transformation property of the solution of the
differential equation under $SL(2,C)$. 
In the calculation of the second order such a technique is not
sufficient and we need a non invertible transformation 
which at the infinitesimal level is related to the operator
$l_{-2}=\frac{1}{z}\frac{\partial}{\partial z}$. Contrary to the $SL(2,C)$
transformations this is not one-to-one in the complex plane. 

On the other hand the procedure we shall describe, involves only the
solutions along the real $z$ axis for $z\geq 1$, and there for $|x|<1$
the transformation is well defined. 

After developing such tools we give the complete second order
computation for the accessory parameter. In so doing we employ a
formalism apt to be extended to higher order computation.

The second order result agrees with the one obtained in 
\cite{ferraripiatek,LLNZ}
by considering the classical limit of the quantum conformal
blocks combined with the exponentiation hypothesis and thus it lends
a strong support to the exponentiation hypothesis of the conformal
blocks in the $b\rightarrow 0$ limit.
We discuss also the extension of the procedure to higher orders.

Obviously, as the determination of the accessory parameter $C(x)$ is
always obtained through the solution of an implicit equation, the fact
that the function $Q(z)$ which represents the energy momentum tensor,
has radius of convergence $1$ in $x$, for $z>1$, does not assure that
the expansion of $C(x)$ in $x$ has the same radius of convergence. For
achieving rigorous lower bounds on such a radius of convergence,
methods similar to those developed in \cite{torusIII,torusI} for the
convergence in the coupling strength should be applied. The developed
technique can also be applied to the problem of the punctured torus.

\section{General setting}

The ordinary differential equation associated with the monodromy problem
is 
\begin{equation}\label{diffequation}
y''(z) + Q(z) y(z)=0
\end{equation}
with 
\begin{equation}
Q(z)= \frac{\delta_0}{z^2}+\frac{\delta}{(z-x)^2}+\frac{\delta_1}{(z-1)^2}
+\frac{\delta_\infty-\delta_0-\delta-\delta_1}{z(z-1)}+\frac{C(x)}{z(z-x)(1-z)}
\end{equation}
where $\delta_j = (1-\lambda_j^2)/4$. $C(x)$ is the accessory
parameter to be fixed so that the monodromy along a contour encircling
both $0$ and $x$ has trace $-2 \cos \pi\lambda_\nu$ and as such it
will depend both on $x$ and $\delta_\nu$. 
We have 
\begin{equation}\label{C0}
C(0) =\delta_\nu-\delta_0-\delta
\end{equation}
and $C(x)$ is related to 
to one used in \cite{ZZ,LLNZ} which we call $C_L(x)$, by 
$C(x) = x(1-x)C_L(x)$ and thus $C(0) =x C_L(x)|_{x=0}$ and 
$C'(0) = [x C_L(x)]'|_{x=0} -C(0)$. 

Expanding in $x$ we have
\begin{equation}
Q=Q_0+x Q_1+ x^2 Q_2 +O(x^3)
\end{equation}
\begin{eqnarray}\label{Qexpansion}
Q_0&=&
\frac{\delta_\nu}{z^2}+\frac{\delta_1}{(z-1)^2}+
\frac{\delta_\infty-\delta_\nu-\delta_1}{z(z-1)}\\
Q_1&=&\frac{2\delta-C'(0)}{z^2(z-1)}-\frac{2\delta+C(0)}{z^3(z-1))}\nonumber\\
Q_2&=& -\frac{C''(0)}{2z^2(z-1)}+\frac{3\delta-C'(0)}{z^3(z-1)}-
\frac{3\delta+C(0)}{z^4(z-1)}.\nonumber
\end{eqnarray}
It is our interest to compute the class of the monodromy along a circuit
enclosing both the origin and $x$. Working near the origin is
difficult due to the singular nature of the kernel. Instead we shall compute
the same monodromy along the circuit shown in fig.1. The great advantage in
performing such a change in the contour is the fact that the expansion in
$x$ of $Q(z)$ along the contour is no longer singular and actually is 
convergent with convergence radius $1$.
\bigskip\bigskip\bigskip
\begin{figure}[htb]
\begin{center}
\includegraphics{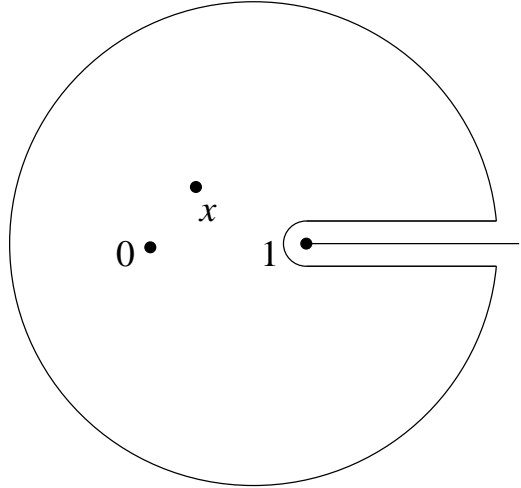}
\end{center}
\caption{the integration contour}
\end{figure}
$Y$ will denote the complex
\begin{equation}
Y(z)= 
\begin{pmatrix}
y_1(z)\\
y_2(z)
\end{pmatrix}
\end{equation}
being $y_k$ two independent solutions of $y_k''+Q_0y_k=0$,
canonical at $z=1$. They are given by
\begin{equation}
y_1(z)=(1-z)^{\frac{1-\lambda_1}{2}}~z^{\frac{1-\lambda_\nu}{2}}
F(\frac{1-\lambda_1-\lambda_\infty-\lambda_\nu}{2},
\frac{1-\lambda_1+\lambda_\infty-\lambda_\nu}{2},1-\lambda_1;1-z)
\end{equation}
\begin{equation}
y_2(z)=(1-z)^{\frac{1+\lambda_1}{2}}~z^{\frac{1+\lambda_\nu}{2}}
F(\frac{1+\lambda_1+\lambda_\infty+\lambda_\nu}{2},
\frac{1+\lambda_1-\lambda_\infty+\lambda_\nu}{2},1+\lambda_1;1-z)~.
\end{equation}
The constant Wronskian is easily computed at $z=1$
\begin{equation}
w_{12}=y_1 y_2'-y'_1 y_2=-\lambda_1~.
\end{equation}

The unperturbed monodromy is computed as follows. 
We start from the $Y$ for real $z$, $z<1$. The continuation to the
upper side of the cut $(1,+\infty)$ is given by
\begin{eqnarray}\label{y1t1+}
y^+_1(z)&=& -i e^{\frac{i\pi\lambda_1}{2}}
(z-1)^{\frac{1-\lambda_1}{2}}~z^{\frac{1-\lambda_\nu}{2}}
F(\frac{1-\lambda_1-\lambda_\infty-\lambda_\nu}{2},
\frac{1-\lambda_1+\lambda_\infty-\lambda_\nu}{2},1-\lambda_1;1-z)\nonumber\\
&\equiv&-i e^{\frac{i\pi\lambda_1}{2}} t_1(z)
\end{eqnarray}
\begin{eqnarray}\label{y2t2+}
y^+_2(z)&=&-i
e^{-\frac{i\pi\lambda_1}{2}}(z-1)^{\frac{1+\lambda_1}{2}}~
z^{\frac{1+\lambda_\nu}{2}}
F(\frac{1+\lambda_1+\lambda_\infty+\lambda_\nu}{2},
\frac{1+\lambda_1-\lambda_\infty+\lambda_\nu}{2},1+\lambda_1;1-z)\nonumber\\
&\equiv&-i e^{-\frac{i\pi\lambda_1}{2}}t_2(z)
\end{eqnarray}
whose asymptotic behavior for large $z$ is
\begin{equation}\label{Y+}
Y^+(z) \approx B^+ 
\begin{pmatrix}
z^{\frac{1-\lambda_\infty}{2}}\\
z^\frac{1+\lambda_\infty}{2}
\end{pmatrix}
=-i
\begin{pmatrix}
B^{(1)}_2 e^\frac{i\pi\lambda_1}{2}&B^{(1)}_1
e^\frac{i\pi\lambda_1}{2}\\
B^{(2)}_1 e^{-\frac{i\pi\lambda_1}{2}}&B^{(2)}_2 e^{-\frac{i\pi\lambda_1}{2}}
\end{pmatrix}
\begin{pmatrix}
z^{\frac{1-\lambda_\infty}{2}}\\
z^\frac{1+\lambda_\infty}{2}
\end{pmatrix}
\end{equation}
$B^{(j)}_{~~k}$ being a well known matrix
\begin{equation}
B^{(j)}_{~~k}= 
\begin{pmatrix}
\frac{\Gamma(1-\lambda_1)\Gamma(\lambda_\infty)}
{\Gamma(\frac{1-\lambda_1+\lambda_\infty-\lambda_\nu}{2})
\Gamma(\frac{1-\lambda_1+\lambda_\infty+\lambda_\nu}{2})} & 
\frac{\Gamma(1-\lambda_1)\Gamma(-\lambda_\infty)}
{\Gamma(\frac{1-\lambda_1-\lambda_\infty-\lambda_\nu}{2})
\Gamma(\frac{1-\lambda_1-\lambda_\infty+\lambda_\nu}{2})}\\
\frac{\Gamma(1+\lambda_1)\Gamma(-\lambda_\infty)}
{\Gamma(\frac{1+\lambda_1-\lambda_\infty+\lambda_\nu}{2})
\Gamma(\frac{1+\lambda_1-\lambda_\infty-\lambda_\nu}{2})} & 
\frac{\Gamma(1+\lambda_1)\Gamma(\lambda_\infty)}
{\Gamma(\frac{1+\lambda_1+\lambda_\infty+\lambda_\nu}{2})
\Gamma(\frac{1+\lambda_1+\lambda_\infty-\lambda_\nu}{2})}
\end{pmatrix}.
\end{equation}
Similarly
\begin{equation}\label{y1t1y2t2-}
y^-_1(z) = i e^{-\frac{i\pi\lambda_1}{2}} t_1(z),~~~~
y^-_2(z) = i e^{\frac{i\pi\lambda_1}{2}}t_2(z)
\end{equation}
and
\begin{equation}\label{Y-}
Y^-(z) \approx B^- 
\begin{pmatrix}
z^{\frac{1-\lambda_\infty}{2}}\\
z^\frac{1+\lambda_\infty}{2}
\end{pmatrix}
=i
\begin{pmatrix}
B^{(1)}_2 e^{-\frac{i\pi\lambda_1}{2}}&B^{(1)}_1
e^{-\frac{i\pi\lambda_1}{2}}\\
B^{(2)}_1 e^{\frac{i\pi\lambda_1}{2}}&B^{(2)}_2 e^{\frac{i\pi\lambda_1}{2}}
\end{pmatrix}
\begin{pmatrix}
z^{\frac{1-\lambda_\infty}{2}}\\
z^\frac{1+\lambda_\infty}{2}
\end{pmatrix}.
\end{equation}
We start from $z=+\infty-i\varepsilon$ eq.(\ref{Y-}), whose continuation
to the upper side of the cut is eq.(\ref{Y+}).
Taking the turn of $2\pi$ at infinity we go back to
$z=+\infty-i\varepsilon$ having encircled the origin and $x$ and we 
obtain the monodromy matrix to lowest order
\begin{equation}
M^0=-(B^+)
\begin{pmatrix}
e^{-i\pi \lambda_\infty}&0\\
0&e^{i\pi \lambda_\infty}
\end{pmatrix}
(B^-)^{-1} .
\end{equation}
One easily checks that ${\rm tr}M^0=-2\cos\pi\lambda_\nu$.
The first order corrections to $Y$ is provided by
\begin{equation}\label{firstiteration}
Y(z)+x S_1(z) Y(z)
\end{equation}
with
\begin{equation}
S_1(z)=\frac{1}{w_{12}}
\begin{pmatrix}
\int_1^z y_2 Q_1 y_1dz&-\int_1^z y_1 Q_1 y_1dz\\
\int_1^z y_2 Q_1 y_2dz&-\int_1^z y_1 Q_1 y_2dz
\end{pmatrix}
\end{equation}
$w_{12} = -\lambda_1$.
Following the procedure illustrated above using (\ref{firstiteration})
instead of the unperturbed $Y$ we obtain the new monodromy
matrix $M^0+\delta M$,
\begin{equation}
\delta M=
x~(S_1^+M^0-~M^0S_1^-)
\end{equation}
with $S_1 \equiv S_1(\infty)$,  
and we have
\begin{equation}
{\rm tr} M = -2\cos\pi \lambda_\nu+x~ {\rm tr} (S_1^+-S_1^-)M^0)
\end{equation}
where 
\begin{equation}\label{Delta11}
(S_1^+-S_1^-)_{12}=\frac{2i\sin \pi\lambda_1}{w_{12}}
\int_1^\infty t_1(z) ~Q_1(z) ~t_1(z)~dz\equiv
\frac{2i\sin \pi\lambda_1}{w_{12}} Q_1(1,1)
\end{equation}
\begin{equation}\label{Delta22}
(S_1^+-S_1^-)_{21}=\frac{2 i \sin \pi\lambda_1}{w_{12}}
\int_1^\infty t_2(z) ~Q_1(z) ~t_2(z)~dz~\equiv
\frac{2 i \sin \pi\lambda_1}{w_{12}} Q_1(2,2)~.
\end{equation}
We have due to eqs.(\ref{y1t1+},\ref{y2t2+},\ref{y1t1y2t2-})
\begin{equation}
(S_1^+-S_1^-)_{11}=-(S_1^+-S_1^-)_{22}=0
\end{equation}
and thus
\begin{equation}\label{trdeltaM}
{\rm tr}~\delta M= x[(S_1^+-S_1^+)_{12}~M^0_{21}+
(S_1^+-S_1^-)_{21}~M^0_{12}]
\end{equation}
and to determine $C'(0)$ we must impose ${\rm tr}~\delta M=0$.

The second order result is obtained by iterating once the result with
$Q_1$ and adding also the contribution obtained by replacing in 
(\ref{firstiteration}) $x~Q_1$ with $x^2Q_2$.

\section{First order calculation} 

The computation of the first order result is reasonably simple. We
shall adopt here a formalism which is apt to be extended to the second
and higher order computation.

We are faced to compute the integrals appearing in 
eqs.(\ref{Delta11},\ref{Delta22}) with $Q_1(z)$ given in 
eq.(\ref{Qexpansion}).
More generally we shall compute analytically the indefinite integrals
\begin{equation}\label{generalintegrals}
\int_1^z \frac{t_j(z)t_k(z)}{z^m(z-1)}dz
\end{equation}
for $m\geq 2$ and where $j$ and $k$ take the value $1$ and $2$.

For $m=2,3$ such integrals can be computed by exploiting 
the transformation properties under $SL(2,C)$ of the solutions. 
Let us consider the equation
\begin{equation}\label{Requation}
\tilde t''+R ~\tilde t=0
\end{equation}
\begin{equation}\label{RequationR}
R=\frac{\delta_\nu}{(z-a)^2}+
\frac{\delta_1}{(z-1)^2}+\frac{\delta_\infty-\delta_1-\delta_\nu}
{(z-a)(z-1)}.
\end{equation}
The solutions of eq.(\ref{Requation}) are
\begin{equation}\label{simpleshift}
\tilde t_1(z,a)=(z-1)^{\frac{1-\lambda_1}{2}}(\frac{z-a}{1-a})^{\frac{1-\lambda_\nu}{2}}
F(\frac{1-\lambda_1-\lambda_\infty-\lambda_\nu}{2},
\frac{1-\lambda_1+\lambda_\infty-\lambda_\nu}{2},1-\lambda_1;
\frac{1-z}{1-a})
\end{equation}
and similarly for $\tilde t_2$.
Then writing
\begin{eqnarray}
R&=& R_0 + a R_1 +O(a^2)\nonumber\\
R_0&=&\frac{\delta_\nu}{z^2}+
\frac{\delta_1}{(z-1)^2}+\frac{\delta_\infty-\delta_1-\delta_\nu}{z(z-1)}
=Q_0,~~~~ 
R_1=\frac{\delta_\infty+\delta_\nu-\delta_1}{z^2(z-1)}-
\frac{2\delta_\nu}{z^3(z-1)}
\end{eqnarray}
\begin{equation}
\dot {\tilde t}_k = \frac{\partial {\tilde t}_k}
{\partial a}\bigg|_{a=0}
\end{equation}
using
\begin{equation}
\dot {\tilde t}_k''+R_0~\dot {\tilde t}_k+R_1~t_k=0
\end{equation}
and $\tilde t(z,0)= t_k(z)$ we have
\begin{equation}\label{firstintegral}
\int_1^z t_k~R_1 ~t_j~dz =t_k'~\dot {\tilde t}_j-t_k 
~\dot {\tilde t}_j'\big\vert_1^z=t_j'~\dot {\tilde t}_k-t_j 
~\dot {\tilde t}_k'\big\vert_1^z~.
\end{equation}
We find
\begin{eqnarray}\label{firstintegrals}
R_1(1,1)\equiv\int_1^\infty t_1~R_1 ~t_1~dz &=& 
-\lambda_\infty^2 B^{(1)}_1 B^{(1)}_2\nonumber\\
R_1(2,2)\equiv\int_1^\infty t_2~R_1 ~t_2~dz &=& 
-\lambda_\infty^2 B^{(2)}_1 B^{(2)}_2~.
\end{eqnarray}

On the other hand the change
$\delta_\nu$ into $\delta_\nu-\varepsilon$ induces in $Q_0$ the change
\begin{equation}\label{deltanuchange}
Q_0\rightarrow Q_0+\delta Q_0,~~~~\delta Q_0 = \frac{\varepsilon}{z^2(z-1)}
\end{equation}
leaving the singularities at $z=1$ and at $z=\infty$ unchanged. This
time the related change $\delta t_k$ is simply given by
\begin{equation}
\delta t_k = \frac{2 \varepsilon}{\lambda_\nu} 
\frac{\partial  t_k}{\partial \lambda_\nu}
\end{equation}
and we can again apply eq.(\ref{firstintegral}) replacing 
$\dot {\tilde t}_k$
with $\frac{2}{\lambda_\nu} \frac{\partial  t_k}{\partial\lambda_\nu}$. 
Combining with (\ref{firstintegrals}) it provides us with
the integrals of type (\ref{generalintegrals}) with $m=2$ and with
$m=3$ appearing in the first order computation, in terms of 
hypergeometric functions and derivatives thereof. 

The integrals appearing in (\ref{Delta11},\ref{Delta22}) have the
upper limit infinity, for which the derived formulae (\ref{firstintegral}) 
also hold. As the
asymptotic behavior of the hypergeometric functions are given
\cite{batemanI} by simple powers of $z$ multiplied by gamma functions,
we have that such integrals are expressed in terms of the functions 
$\Gamma$ and $\psi$, where $\psi(x)=\Gamma'(x)/\Gamma(x)$.
We shall denote by $N_m$ the expression
\begin{equation}\label{generalintegrals}
N_m=\frac{1}{z^m(z-1)}.
\end{equation}
They form a basis for the the derivative of $Q(z)$ with respect to $x$
to any order and we shall set
\begin{equation}\label{generalintegrals}
\int_1^\infty t_k\frac{1}{z^m(z-1)}t_jdz = N_m(k,j)~.
\end{equation}
Explicitly we find
\begin{equation}
N_2(1,1)=\frac{\lambda_\infty}{\lambda_\nu}B^{(1)}_1
B^{(1)}_2\Psi(\lambda_1,\lambda_\nu,\lambda_\infty)
\equiv\frac{\lambda_\infty}{\lambda_\nu}B^{(1)}_1
B^{(1)}_2\Psi_1
\end{equation}
and
\begin{eqnarray}
N_2(2,2)=\frac{\lambda_\infty}{\lambda_\nu}B^{(2)}_1
B^{(2)}_2\Psi(-\lambda_1,-\lambda_\nu,-\lambda_\infty)
\equiv\frac{\lambda_\infty}{\lambda_\nu}B^{(2)}_1
B^{(2)}_2\Psi_2
\end{eqnarray}
where we defined
\begin{eqnarray}
\Psi(\lambda_1,\lambda_\nu,\lambda_\infty) &\equiv&
\psi(\frac{1-\lambda_1-\lambda_\infty-\lambda_\nu}{2})-
\psi(\frac{1-\lambda_1+\lambda_\infty-\lambda_\nu}{2})-\nonumber\\
& &\psi(\frac{1-\lambda_1-\lambda_\infty+\lambda_\nu}{2})+
\psi(\frac{1-\lambda_1+\lambda_\infty+\lambda_\nu}{2})~.
\nonumber
\end{eqnarray}
Moreover in eq.(\ref{trdeltaM}) we have
\begin{equation}               
M^0_{12}~~{\rm det}(B^-)=-2~i~\sin\pi\lambda_\infty~~B^{(1)}_1
B^{(1)}_2
\end{equation}
\begin{equation}
M^0_{21}~~{\rm det}(B^-)=~~2~i~\sin\pi\lambda_\infty~~B^{(2)}_1
B^{(2)}_2~.
\end{equation}
with ${\rm det}B^-=\lambda_1/\lambda_\infty$.
For future use we report below also the values of $M^0_{11}$ and $M^0_{22}$.
\begin{equation}
M^0_{11}~~{\rm det}(B^-)=e^{i\pi\lambda_1}(B^{(1)}_1 B^{(2)}_1
e^{i\pi\lambda_\infty}-B^{(1)}_2 B^{(2)}_2
e^{-i\pi\lambda_\infty})
\end{equation}
\begin{equation}
M^0_{22}~~{\rm det}(B^-)=e^{-i\pi\lambda_1}(B^{(1)}_1 B^{(2)}_1
e^{-i\pi\lambda_\infty}-B^{(1)}_2 B^{(2)}_2
e^{i\pi\lambda_\infty})~.
\end{equation}

The vectors
\begin{eqnarray}               
T(1,1)&=& \big(R_1(1,1), N_2(1,1)\big) = 
B^{(1)}_1B^{(1)}_2 \big(-\lambda_\infty^2,
\frac{\lambda_\infty}{\lambda_\nu}\Psi_1\big)
\equiv B^{(1)}_1B^{(1)}_2 \hat T(1,1)\\               
T(2,2)&=& \big(R_1(2,2), N_2(2,2)\big) = B^{(2)}_1B^{(2)}_2 \big(-\lambda_\infty^2,
\frac{\lambda_\infty}{\lambda_\nu}\Psi_2\big)
\equiv B^{(2)}_1B^{(2)}_2 \hat T(2,2)
\end{eqnarray}
are given by
\begin{equation}               
T(k,k) =A
\begin{pmatrix}
N_2(k,k)\\
N_3(k,k) 
\end{pmatrix}
\end{equation}
with
\begin{equation}               
A=
\begin{pmatrix}
\delta_\nu+\delta_\infty-\delta_1&-2\delta_\nu\\
1&0
\end{pmatrix} 
~.
\end{equation}
$Q_1$, see eq.(\ref{Qexpansion}), in the basis $N_2,N_3$ 
is represented by the vector
\begin{equation}               
q_1 = (2\delta-C'(0),-2\delta-C(0))
\end{equation}
and thus, see eq.(\ref{trdeltaM}), the equation for $C'(0)$ becomes
\begin{equation}               
0= q_1\cdot A^{-1}\cdot (\hat T(1,1) -\hat T(2,2)) 
\end{equation}
i.e.
\begin{eqnarray}               
0&=&2\delta_\nu(2\delta-C'(0))-
(2\delta+C(0))(\delta_\nu+\delta_\infty-\delta_1)=\\
& &-2\delta_\nu(C'(0)+C(0))-(-\delta_\nu+\delta_\infty-\delta_1)
(\delta_\nu-\delta_0+\delta)
\end{eqnarray}
giving
\begin{equation}\label{C1}
C'(0)
=\frac{(\delta_\nu-\delta_0+\delta)(\delta_\nu-\delta_\infty+\delta_1)}
{2\delta_\nu}-C(0) = [xC_L(x)]'|_{x=0}-C(0)
\end{equation}
which is the result of \cite{ZZ,ferraripiatek,LLNZ} obtained by taking the
$b\rightarrow 0$ limit of the conformal blocks.

\section{Second order calculation}

The equation which gives $C''(0)$ is provided by the vanishing of the
coefficient of $x^2$ in the expansion of
\begin{equation}
{\rm tr}(1+xS^+_1+x^2S^+_2)M^0(1+xS^-_1+x^2S^-_2)^{-1}
\end{equation}
i.e.
\begin{equation}\label{secondordereq}
0 = {\rm tr}(S^+_2-S^-_2)M^0-{\rm tr}(S^+_1-S^-_1)M^0S^-_1~.
\end{equation}
The second order change in the functions $y_k$ is 
given by the direct contribution
due to $Q_2$ and by the second iteration of the contribution of $Q_1$.

With regard to the direct contribution we have to compute
\begin{equation}
\int_1^\infty t_k(z) Q_2(z)t_j(z)~dz,
\end{equation}
with $Q_2(z)$ given in eq.(\ref{Qexpansion}), 
where the new integrals $N_4(k,j)$ appear. In the basis $N_2,N_3,N_4$ 
$Q_2$ is represented by the vector
\begin{equation}
q_2 = (-C''(0)/2,3\delta-C'(0),-3\delta-C(0))~.
\end{equation}
The $N_4(j,k)$ cannot be
computed by performing an $SL(2,C)$ transformation. We shall exploit the
new transformation
\begin{equation}\label{lm2transf}
z =\frac{v - \frac{a}{v}}{1-a}
\end{equation}
and use the  schwarzian transformation of $R_0$ and the
$-\frac{1}{2}$-form nature \cite{hawley} of the solutions $t_k$. 
The above transformation gives rise to
\begin{equation}
R(v,a) = Q_0\big(\frac{v^2 - a}{v(1-a)}\big)
\big(\frac{dz}{dv}\big)^2 -\{z,v\}
\end{equation}
where
\begin{equation}
\{z,v\}= -\frac{3a}{(a+v^2)^2}
\end{equation}
is the Schwarz derivative of the transformation 
and the new solutions are
\begin{equation}\label{tkva}
t_k(v,a) = \frac{1}{\sqrt{1-a}}\bigg(\frac{dz}{dv}\bigg)^{-\frac{1}{2}}
t_k\bigg(\frac{v^2 - a}{v(1-a)}\bigg)
=\frac{v}{\sqrt{v^2+a}}~
t_k\bigg(\frac{v^2 - a}{v(1-a)}\bigg).
\end{equation}

Reverting to the $z$-notation for the variable and denoting with the
dot the derivative w.r.t. $a$ we have
\begin{equation}
R(z,a) = Q_0\bigg(\frac{z^2 - a}{z(1-a)}\bigg)
\bigg(\frac{z^2+a}{z^2(1-a)}\bigg)^2 +\frac{3a}{(a+z^2)^2}
\end{equation}
with
\begin{equation}\label{dotR}
\dot R(z,0) = \frac{\delta_\nu-\delta_1-\delta_\infty}{z^2(z-1)}
+\frac{3-3\delta_1+3\delta_\infty+\delta_\nu}{z^3(z-1)}
-\frac{3+4\delta_\nu}{z^4(z-1)}~.
\end{equation}
From eq.(\ref{tkva}) we have
\begin{equation}\label{dottva}
\dot t_k(z,0) = -\frac{1}{2z^2}t_k(z)+ (z-\frac{1}{z})~t_k'(z).
\end{equation}
Contrary to the $SL(2,C)$ transformations the transformation 
(\ref{lm2transf}) is not
one-to-one in the complex plane. 
Nevertheless for $|a|<1$ the transformation is well defined, i.e. non
singular along the line $1<z<\infty$ which is our range of application.

We can then exploit again the integration formula
(\ref{firstintegral}), with $\dot {\tilde t}_k$ replaced by
eq.(\ref{dottva}) and $R_1$ by $\dot R$ of eq.(\ref{dotR}).
One can also verify the correctness of the result
by taking explicitly the derivative w.r.t. $z$ of the obtained formula.

We introduce the three dimensional vectors $T^{r}(k,j)$, with
$r=1,2,3$ which represent the matrix elements of the variation of $Q_0$
under respectively the transformation of eq.(\ref{RequationR}) ($r=1$), the
of eq.(\ref{deltanuchange}) ($r=2$) and of the 
transformation of the above eq.(\ref{lm2transf}) ($r=3$).
We have
\begin{equation}\label{Tjk}
T(j,k)= A
\begin{pmatrix}
N_2(j,k)\\
N_3(j,k)\\
N_4(j,k)
\end{pmatrix}
\end{equation}
with
\begin{equation}\label{dottkva}
A=
\begin{pmatrix}
\delta_\infty+\delta_\nu-\delta_1&-2\delta_\nu&0\\
1&0&0\\
\delta_\nu-\delta_1-\delta_\infty&3-3\delta_1+3\delta_\infty+\delta_\nu&
-3-4\delta_\nu
\end{pmatrix}
\end{equation}
and
\begin{eqnarray}\label{Tjk3dim}
T(1,1)&=&B^{(1)}_1
B^{(1)}_2~(-\lambda^2_\infty,\frac{\lambda_\infty}{\lambda_\nu}\Psi_1,
-\lambda^2_\infty)\equiv B^{(1)}_1 B^{(1)}_2 \hat T(1,1)\nonumber\\
T(2,2)&=&B^{(2)}_1 B^{(2)}_2~
(-\lambda^2_\infty,\frac{\lambda_\infty}{\lambda_\nu}\Psi_2,
-\lambda^2_\infty)\equiv B^{(2)}_1 B^{(2)}_2 \hat T(2,2)~.
\end{eqnarray}
The inversion of eq.(\ref{Tjk}) provides the values of the fundamental
matrix elements $N_m(j,k)$, $m=2,3,4$. We notice that the procedure
can be extended to all values of $m$ by considering the variation of
the equation $y''+R_0y=0$ under the transformation 
\begin{equation}
z= (v-\frac{a}{v^{m-3}})/(1-a)~.
\end{equation}
In addition due to the structure of the matrices $A$
such a procedure is purely iterative, i.e. known $N_m(j,k)$ for
$m=2,3,\dots,n$ the computation of $T^n(j,k)$, provides
directly $N_{n+1}(j,k)$.

In the present case we have
\begin{equation}
Q_2(j,k) = q_2\cdot A^{-1}\cdot T(j,k)~.
\end{equation}

We come now to the second iteration of $Q_1$. Given the Green function
\begin{equation}
G(z,z')=\frac{1}{w_{12}}(y_1(z)y_2(z')-y_2(z)y_1(z'))
\end{equation}
the expression we are confronted with, for the second order
change of $y_k(z)$ is
\begin{equation}\label{seconditeration}
y_1(z)\frac{1}{w_{12}}\int_1^z Q_1(z') y_2(z')\delta^{(1)} y_k(z')dz'-
y_2(z)\frac{1}{w_{12}}\int_1^z Q_1(z') y_1(z')\delta^{(1)} y_k(z')dz'
\end{equation}
with $\delta^{(1)} y_k$ given by eq.(\ref{firstiteration}).
The indefinite integrals appearing in (\ref{seconditeration}) can be
obtained from the equation
\begin{equation}
\tilde y''(z,a)+\tilde Q(z,a)\tilde y(z,a)=0
\end{equation} with
\begin{equation}
\tilde Q(z,a)=\frac{\delta_\nu+ s a}{(z- c a)^2}+\frac{\delta_1}{(z-1)^2}+
\frac{\delta_\infty-\delta_1-\delta_\nu - s a}{(z- c a)(z-1)}
\end{equation}
where we shall impose 
\begin{equation}\label{newQ1}
\frac{\partial \tilde Q(z,a)}{\partial a}\bigg|_{a=0}=
\frac{2\delta-C'(0)}{z^2(z-1)}-\frac{2\delta+C(0)}{z^3(z-1)}=Q_1(z)
\end{equation}
with $C(0)= \delta_\nu-\delta_0-\delta$ and $C'(0)$ given by eq.(\ref{C1}).
In order to fit the coefficients of $N_2,N_3$ in $Q_1$ we have to
allow in principle for two parameters $c$ and $s$. We find
\begin{equation}\label{c}
c = \frac{\delta+\delta_\nu-\delta_0}{2\delta_\nu},~~~~s =0~.
\end{equation}

Notice that the transformation leading from $Q_0(z)$ to
$\tilde Q(z,a)$ is of the same type as the one appearing in 
(\ref{Requation},\ref{RequationR}) 
of which we know the solutions. It follows that after replacing 
$C(0)$ and $C'(0)$ in (\ref{newQ1}) with their values,
the simplest method to
compute  the matrix elements $Q_1(j,k)$ is to use the transformation
(\ref{simpleshift}) with $a$ replaced by $c a$. 
One can easily prove that 
\begin{equation}\label{correctedsolution}
y_1(z)+ x\frac{\partial \tilde y_1(z,a)}{\partial a}\bigg|_{a=0}
\equiv y_1(z)+ x\dot{\tilde y}_1(z)
\end{equation}
has the correct boundary condition 
$(1-z)^{(1-\lambda_1)/2}$ with coefficient $1$ at $z=1$, as imposed by the
solution of the Volterra equation 
and thus $x\dot{\tilde y}_1(z)$ equals $\delta^{(1)}y_1$ i.e.
it is the first order correction to $y_1$. It is expressed in
terms of derivatives of the hypergeometric function. 
The same holds for $y_2$.
Then the integrals
\begin{equation}
\int_1^z y_k(z') Q_1(z') \delta^{(1)} y_j(z') dz' 
\end{equation}
appearing in (\ref{seconditeration}) can be computed as follows. Using
\begin{equation}
\tilde Q(z,a)=Q_0(z)+a Q_1(z)+a^2\tilde Q_2(z)+\cdots 
\end{equation} 
and
\begin{equation}
\ddot{\tilde y}_k''+2 \tilde Q_2(z)y_k(z)+2Q_1(z)\dot{\tilde{y}}_k(z)+
Q_0(z)\ddot{\tilde{y}}_k(z)=0
\end{equation}
we have
\begin{equation}\label{secondintegral}
\int_1^zy_k(z')Q_1(z')\dot{\tilde y}_j(z')dz'
 =-\frac{1}{2}(y_k(z)\ddot{\tilde y}_j'(z)-y_k'(z)\ddot{\tilde y}_j(z))\big|_1^z
-\int_1^z y_k(z') \tilde Q_2(z') y_j(z') dz'~.
\end{equation}

Notice that $\tilde Q_2$ is not equal to the $Q_2$ of eq.(\ref{Qexpansion})
but it will be expedient for computing the l.h.s. of
(\ref{secondintegral}). $\tilde Q_2$ in the base $N_2,N_3,N_4$,  
is represented by the vector
\begin{equation}
\tilde q_2=c^2~(0,\delta_\infty-\delta_1+2\delta_\nu,-3\delta_\nu)~.
\end{equation}
 
We know $\ddot{\tilde y}_k(z)$ and $\ddot{\tilde y}_l'(z)$ 
and as a result we know the l.h.s. of eq.(\ref{secondintegral})
thus providing the second iteration of the Volterra equation
in terms of hypergeometric functions and derivatives thereof.
Explicitly we find in eq.(\ref{secondordereq})
\begin{eqnarray}\label{puresecondorder}
& &{\rm tr}(S^+_2-S^-_2)M^0=\\
& &4 \frac{\sin \pi\lambda_1}{\lambda_1}\sin\pi\lambda_\infty
B^{(1)}_1B^{(1)}_2B^{(2)}_1B^{(2)}_2
\big\{(q_2-\tilde q_2)\cdot A^{-1}\cdot (\hat T(1,1)-\hat T(2,2))
-\lambda_\infty^2\lambda_1c^2\big\}~.\nonumber
\end{eqnarray}
The computation of the term $-{\rm tr}(S^+_1-S^-_1)M^0S^-_1$ in 
eq.(\ref{secondordereq})
requires simply the knowledge of $S^\pm_1$ which we have already computed
and it cancels the term $-\lambda_\infty^2\lambda_1c^2$ in the curly
brackets in the above equation. To summarize the equation for
$C''(0)$ is given by
\begin{equation}
0=(q_2-\tilde q_2)\cdot A^{-1}\cdot(\hat T(1,1)-\hat T(2,2))~.
\end{equation}
Due to the structure of $\hat T(j,k)$, see eq.(\ref{Tjk3dim}), the vector 
$\hat T(1,1)-\hat T(2,2)$ has a single entry different from zero
and we have for $C''(0)\equiv [x C_L(x)]''_{x=0}-2 C'(0)-2 C(0)$ 
\begin{eqnarray}
& &C''(0)=-\frac{(\delta_\infty+\delta_\nu-\delta_1)
[C'(0)-3\delta+c^2(2\delta_\nu+\delta_\infty-\delta_1)]}{\delta_\nu}\\
&-&\frac{(C(0)+ 3\delta-3c^2\delta_\nu)
[3\delta_1^2+3\delta_\nu^2+3\delta_\infty(1+\delta_\infty)+
\delta_\nu(3+2\delta_\infty)-3\delta_1(1+2\delta_\nu+2\delta_\infty)]}
{\delta_\nu(3+4\delta_\nu)}\nonumber
\end{eqnarray}
where $C(0)$, $C'(0)$ and $c$ are given respectively by 
eqs.(\ref{C0},\ref{C1},\ref{c}). The value of $C''(0)$ agrees with the one
obtained in \cite{ferraripiatek} and \cite{LLNZ} by taking the 
$b\rightarrow 0$ limit of the
conformal blocks thus providing strong support to the exponentiation 
hypothesis.

The procedure can also be pushed to higher order even if we need a
systematic organization of the mixed contributions.

Integrals similar to those discussed here appear in the accessory
parameter problem for the torus, dealt with in
\cite{hyperbolic}. There the logarithmic part was derived to all
orders and compared with success with the saddle point prediction on
the quantum theory, while the presence of integrals of the above
mentioned type hampered the analytical evaluation of the $q$ term of
the expansion, $q$ being the nome of the torus. Now we have the
possibility of computing analytically not only the $q$ term but also
the $q^2$ term.

\section{Conclusions}

In this paper we developed, for the monodromy problem considered in
papers \cite{LLNZ,LRS} an analytical technique to compute the expansion of
the accessory parameter in term of the invariant cross ratio directly,
without taking the limit of the quantum conformal blocks for infinite
central charge. In the first order computation it is shown how the 
integrals containing products of two hypergeometric functions, 
which appear in the first iteration of the Volterra equation can be 
computed analytically in terms of $\Gamma$-functions and derivatives
thereof, i.e. $\psi$-functions. The method is to exploit
the transformation properties of the kernel under $SL(2C)$ and the
$-1/2$-form nature of the solutions. 

The computation to second order involves double-integrals of products
of four hypergeometric functions, and we show how also these can be
computed analytically. In such second order computation we need a
transformation which at the infinitesimal level is akin to the
$L_{-2}$ generator of the Virasoro algebra.  The transformation is not
one-to-one on the complex plane but is well defined in the region
needed for our computations.

Both our first and second order results agree with the ones obtained
from the classical limit of the quantum conformal blocks under the
exponentiation hypothesis \cite{ferraripiatek,LLNZ}
and thus they lend strong support to such
exponentiation hypothesis in the classical  $b\rightarrow 0$ limit.

With regard to the extension to higher orders we have shown how the
fundamental matrix elements $N_m(j,k)$ can all be computed by a simple
iterative procedure. As it happens already to second order in the
$n$-th order computation one has both direct contribution from the
$n$- order term in the expansion of the kernel and mixed contributions
due to lower order expansion. One should need a systematic
organization of such mixed contribution to go to arbitrary order.

In the full treatment of the torus with one source \cite{hyperbolic},
integrals of product of hypergeometric function, similar to those
which we discussed above, appeared. The possibility of computing them
analytically will allow an extension of the results reported in
\cite{hyperbolic}.

\vfill

\end{document}